# Changing User Attitudes to Reduce Spreadsheet Risk

Dermot Balson

Perth, Australia

Dermot.Balson@Gmail.com

**ABSTRACT**

*A business case study on how three simple guidelines:*

*1. make it easy to check (and maintain)*
*2. make it safe to use*
*3. keep business logic out of code*

*changed user attitudes and improved spreadsheet quality in a financial services organisation.*

## 1 INTRODUCTION

A few years ago, my former employer, a large financial services organisation, became concerned about the impact of poor skill levels and faulty spreadsheets, and so it researched what could be done to reduce risk of error, a process that took several months.

This exercise was fairly small scale, in that it affected fewer than 100 employees within a particular division of the organisation.

The results were surprising, because they did not primarily point to additional training and discipline, as expected. Instead, the biggest driver of spreadsheet quality was found to be user attitudes.

## 2 THE RESEARCH

The following is only a summary of the research, because the business objective was to find a solution as quickly as possible, so there were no formal records.

It was difficult to find a single comprehensive source for what was wrong with spreadsheets and how to fix it. We found the many papers offered by Eusprig to be very useful, and Ray Panko's research into spreadsheet errors was particularly valuable.

We also researched many other sources, especially in the fields of software programming and, in particular, software testing, where a great deal of effort has gone into understanding and managing errors. The error rates and patterns were very similar to those for spreadsheets.

Finally, we looked at error prevention in industries where mistakes can be fatal, such as airlines, space missions and nuclear power.

From all of this, we learned that

- everybody makes mistakes, even experts [1]
- spreadsheet mistakes vary widely [1,2]
- mistakes happen even on large, important models built by finance 'experts' [4]
- models need to be checked, but it is almost impossible to eliminate errors
- checking is hard to do well, and requires a certain mindset [3,5]

Once we realised that errors had to be managed because they could not be eliminated, and that checking was difficult, we had a good look at our own organisation, and it came as quite a shock.

Our IT department potentially had the professional training needed to control spreadsheet use, but, as in many organisations, its focus was on maintaining a complex system, its skills were in networking rather than in programming, and it had little interest in what users were doing with spreadsheets.

Our business managers similarly were focussed on day to day issues, and there was little supervision of spreadsheet use.

The result was generally that users taught themselves, or learned from the people they worked with. Some formal training was available, but it tended to be functional (eg this is how to use a pivot table), and there was no training in how to work safely and minimise error.

To make it worse, most heavy spreadsheet users were relatively young and did not think too hard about risk of error, and, even where users did develop reasonable skills, some of them did not use them safely, instead building complex and unmanageable models.

Finally, we had tried to use spreadsheet guidelines before, but giving users lists of do's and don'ts had been ineffective.

It became very clear that we had to completely change the way we managed spreadsheets, and not just focus on user skills.

**3 SAFETY GUIDELINES**

Fortunately, as we had many mining clients, we had seen how they managed safety, and it was as much to do with office culture as with formal training. For example, meetings might always start with a safety tip, or there might be a rule to always use the handrail on every staircase. You could not escape the safety message, wherever you went.

So while we did plan additional training, we decided to try attitudinal change as our key defence against error, keeping it as simple as possible.

The result was the following three guidelines. In each case, we focussed on getting the message across first, and then showing users techniques for applying the guidelines.

### 3.1 Make it easy to check (and maintain)

We knew that checking couldn't find all the errors. It was going to be harder if the checker was confused by the layout, couldn't read pink text on a green background, or was baffled by a formula with 26 functions in it.

So we asked users to give the checker the best chance to find faults, by making models easy to check. This includes laying models out clearly, documenting them fully, keeping them simple, not being clever, and giving people what they expected to see, so they didn't waste time figuring out what's going on.

An interesting side effect was that it helped avoid style wars, such as whether range names should be used everywhere or nowhere. The guideline simply asked what was best for this particular spreadsheet, and - importantly – what was best for your checker(s), rather than yourself. As we told users "you build spreadsheets for other people, not yourself".

The guideline also encouraged the use of common standards. If, for example, I am deciding whether to use range names, the answer may vary for different models. But I may always colour input cells green, because that is a standard in my workgroup.

We also suggested including self-checks and reasonableness tests which exposed errors, including where possible, the tests that a checker would make. As part of the attitudinal change, we put responsibility on users for proving that their spreadsheets were right as far as possible, rather than relying on the checker to find any mistakes.

Maintenance was also important if a model was in regular use, and so we asked users to think about what would make it easier for a new user to make changes safely to the model.

This particular guideline was fairly easy to sell, because everyone had had the experience of opening up incomprehensible old spreadsheets, and it was hard to argue against making life easier for your colleagues, especially when you might have to check their spreadsheets!

### 3.2 Make it safe to use

Most models were used by at least two people, so there was the risk of misunderstandings and misuse. So getting the model right was only the first step – making it safe to use was just as important.

There are many ways to do this. Inputs are safer if grouped together and clearly labelled, explained and validated. Key formulae and results are safer if protected from change. Reports should be complete and unambiguous.

We didn't specify how far this should be taken, relying on common sense to judge how much effort was justified for any particular model.

If a model was going to be in use for a while, we suggested the test of whether someone new to the model could follow it clearly without any verbal explanation.

### 3.3 Keep all business logic out of code

We had some models which included VBA code, and all too often, they were very hard to maintain, especially if the author had left. Also, very few users could read or safely change VBA code.

We found that most code changes were to do with business logic, ie things like commission rates or sales forecasts, and that generic automation code rarely changed.

Accordingly, we added a third guideline, to keep all business logic out of code wherever possible. This meant that while code was fine for generic automation and repetitive tasks, business formulae should be kept in worksheets.

This had a twofold result, not only making the business logic accessible to all users, but also making our code easier to maintain, and also more generic, allowing it to be shared across different spreadsheets.

### 3.4 Management support

An attitude change requires management support because office culture is usually driven from the top down. The advantage of the guidelines above was that they were extremely simple and even non-technical managers could see their value. But if they were not strongly supported and embedded in the work culture, then little was going to change.

### 4 DID THEY WORK?

We had mixed results in implementing the guidelines. Users were scattered across several cities, and the enthusiasm of local management varied because they had many other priorities.

However, in the workgroups where the guidelines were actively adopted, there was a considerable improvement in quality, mainly because users were actively thinking about safety. We saw cleaner and simpler models, much better documentation and testing, and greater willingness to learn new techniques.

We would have liked to support this with more skills training, but external training tended to be ineffective, and internal training was expensive because it tied up key users (who weren't necessarily good teachers, either).

We found that the guidelines gave us a non-confrontational way of criticising spreadsheets, because the focus was always on how other users would see those spreadsheets, rather than the original developer. This made criticism less personal, and made it easier to get rid of annoying personal preferences such as bright purple or bolding everywhere.

What surprised us was that some of the biggest attitude changes occurred in the advanced users, who we expected to be resistant to anything that would restrict their ability to create models freely. Instead, they largely regarded the guidelines as an additional challenge which gave them an even greater opportunity to use their skills. This meant that even a relatively boring model could be made interesting by making it as simple as possible.

**5 CONCLUSIONS**

This experience suggests that user attitudes can be effectively addressed with attitudinal guidelines, provided they are strongly supported by management and reinforced with skills training.

Guidelines are not a substitute for training and discipline, but they provide relevance for safety techniques, and they also appear to make users more receptive to improving their skills, leading to more effective training.